\documentstyle[epsf]{article} 

\begin{document}            

\renewcommand\baselinestretch{1.3}
\large\normalsize

\title{Local magnetic properties
      \protect\\ 
      of periodic nonuniform spin-$\frac{1}{2}$ $XX$ chains}

\author{
Oleg Derzhko$^{1,2}$,
Johannes Richter$^{3}$
and
Oles' Zaburannyi$^{1}$\\
\small{$^1$Institute for Condensed Matter Physics,}\\
\small{1 Svientsitskii Street, L'viv--11, 79011, Ukraine}\\
\small{$^2$Chair of Theoretical Physics, 
Ivan Franko National University in L'viv,}\\
\small{12 Drahomanov Street, L'viv--5, 79005, Ukraine}\\
\small{$^3$Institut f\"{u}r Theoretische Physik,
Universit\"{a}t Magdeburg,}\\
\small{P.O. Box 4120, D--39016 Magdeburg, Germany}}

\date{\today}

\maketitle

\begin{abstract}
Using the Jordan--Wigner fermionization, 
Green function approach 
and continued fractions 
we examine rigorously  
the local magnetizations 
and 
the local static susceptibilities  
of the spin-$\frac{1}{2}$ $XX$ chain in a transverse field
with regularly varying exchange interactions. 
We discuss our findings from a viewpoint
of the strong--coupling approach.
\end{abstract}

\vspace{3mm}

 
\vspace{3mm}


\vspace{5mm}

\noindent
{\bf {Postal addresses:}}\\

\vspace{0mm}

\noindent
Dr. Oleg Derzhko (corresponding author)\\
Oles' Zaburannyi\\
Institute for Condensed Matter Physics\\
1 Svientsitskii Street, L'viv-11, 79011, Ukraine\\
tel/fax: (0322) 76 19 78\\
email: derzhko@icmp.lviv.ua\\

\vspace{0mm}

\noindent
Prof. Johannes Richter\\
Institut f\"{u}r Theoretische Physik,
Universit\"{a}t Magdeburg\\
P.O. Box 4120, D-39016 Magdeburg, Germany\\
tel: (0049) 391 671 8841\\
fax: (0049) 391 671 1217\\
email: Johannes.Richter@Physik.Uni-Magdeburg.DE

\clearpage

\renewcommand\baselinestretch{1.9}
\large\normalsize

\section{Introduction}

In recent years there has been tremendous interest  
in the spin models of one--dimensional magnets  
because materials sciences have made a remarkable progress 
and a number of quasi--one--dimensional magnets 
have become available. 
On the one hand, 
the one--dimensional magnets 
may exhibit a wide variety of peculiar properties
(e.g., gapless or gapped low--energy excitations,
magnetization plateaus, spin--Peierls phases etc.).
On the other hand, 
for the one--dimensional spin models 
specific theoretical tools can be applied 
(Jordan--Wigner fermionization,
bosonization,
Bethe ansatz etc.) 
and therefore a more comprehensive analysis can be performed.
Although usually the Heisenberg model with modification is used 
to describe the properties 
of low--dimensional magnetic compounds, 
some of their generic features
can be illustrated within the simpler framework
of the $XX$ model in a transverse field.

In the present paper 
we study the effects of regular modulation of the exchange interactions
on the magnetization and static susceptibility 
examining for this purpose
the local magnetic properties 
of the spin-$\frac{1}{2}$ $XX$ chain in a transverse field. 
A model with the alternating exchange interactions arises 
while describing superlattices \cite{001}
or annealed ``bond impurities'' \cite{002}. 
Recent study 
of the alternating quantum (Heisenberg) spin chain 
with experimental applications 
has been reported in \cite{003}.
Quantum spin chains with regularly modulated exchange interactions 
attract much attention because of the magnetization plateaus 
which can be realized in such chains.
Oshikawa, Yamanaka and Affleck \cite{004}
considered a general quantum spin chain  
exhibiting rotational symmetry 
with respect to the direction of the applied uniform field
and proposed the necessary condition for magnetization plateaus 
as $p(s-m)={\mbox{integer}}$,
where $p$ is the period of chain, 
$s$ is the spin value,
$m$ is the zero temperature magnetization per site.
Much work has been done
[5--9]
to investigate magnetization plateaus 
in various quantum spin chains
using approximate analytical approaches
(mainly, bosonization plus renormalization group analysis
or 
strong--coupling limit plus perturbation expansions)
and numerical ones
(mainly, exact diagonalization
or
density--matrix renormalization group 
techniques). 
On the other hand, 
for a simpler spin-$\frac{1}{2}$ $XX$ chain 
the magnetization profiles can be calculated exactly 
for any finite period of nonuniformity $p$ 
\cite{010}. 
Moreover,
as we shall show below, 
it is also possible to examine rigorously 
a nonuniversal behavior 
of the {\em local magnetizations} of such chains. 
In addition, we can calculate 
the {\em local static susceptibilities}.
The local magnetic properties of quantum spin chains 
can be probed experimentally. 
Thus, 
the precise analysis of the NMR line shapes 
can yield the on--site magnetizations 
in the corresponding quasi--one--dimensional compounds 
\cite{011,012}.

The paper is organized as follows.
First we explain 
how the local magnetic properties 
of the regularly nonuniform 
spin-$\frac{1}{2}$ $XX$ chain in a transverse field 
can be calculated exactly. 
Then we analyze the obtained results 
for the zero temperature magnetization profiles 
from a viewpoint of the strong--coupling limit. 
We show to what extent the exact zero temperature magnetization profiles 
can be reproduced 
within the framework of the strong--coupling approximation.
Further 
we discuss the temperature dependence of local static susceptibilities.
We end up with a brief summary.

\section{Continued fractions 
and the local magnetic properties} 

Consider 
a nonuniform 
spin-$\frac{1}{2}$ $XX$ chain 
in a transverse field
defined by the Hamiltonian
\begin{eqnarray}
\label{001}
H=\sum_{n=1}^N\Omega_ns_n^z
+2\sum_{n=1}^NI_n
\left(s_n^xs_{n+1}^x+s_n^ys_{n+1}^y\right)
\\
=\sum_{n=1}^N\Omega_n\left(s_n^+s_n^--\frac{1}{2}\right)
+\sum_{n=1}^NI_n
\left(s_n^+s_{n+1}^-+s_n^-s_{n+1}^+\right).
\nonumber
\end{eqnarray}
By applying the Jordan--Wigner transformation 
one comes to spinless fermions with the Hamiltonian
\begin{eqnarray}
\label{002}
H=\sum_{n=1}^N\Omega_n\left(c_n^+c_n-\frac{1}{2}\right)
+\sum_{n=1}^NI_n
\left(c_n^+c_{n+1}-c_n c_{n+1}^+\right).
\end{eqnarray}
We shall be concerned with the local on--site magnetization  
$m_n=\langle s_n^z\rangle
=\langle s_n^+s_n^-\rangle-\frac{1}{2}
=\langle c_n^+c_n\rangle-\frac{1}{2}$
and the local on--site static susceptibility  
$\chi_n=\frac{\partial m_n}{\partial\Omega}$.
Here the angular brackets denote the canonical thermodynamic average.

Let us introduce the Green functions \cite{013}
$G_{nm}^{\mp}(t)=\mp{\mbox{i}}\theta(\pm t)
\langle\left\{ c_n(t), c_m^+\right\}\rangle,$ 
$G_{nm}^{\mp}(t)=\frac{1}{2\pi}
\int_{-\infty}^{\infty}{\mbox{d}}\omega
{\mbox{e}}^{-{\mbox{i}}\omega t}G_{nm}^{\mp}
(\omega\pm{\mbox{i}}\epsilon),$
$\epsilon\rightarrow+0,$
and note 
that for the model defined by Eq. (\ref{002})
the following continued fraction representation 
for $G_{nn}^{\mp}\equiv G_{nn}^{\mp}(\omega\pm{\mbox{i}}\epsilon)$
holds
\begin{eqnarray}
\label{003}
G_{nn}^{\mp}
=\frac{1}{\omega\pm{\mbox{i}}\epsilon-\Omega_n
-\Delta_n^--\Delta_n^+},
\\
\Delta_n^-
=\frac{I_{n-1}^2}{\omega\pm{\mbox{i}}\epsilon-\Omega_{n-1}
-\frac{I_{n-2}^2}{\omega\pm{\mbox{i}}\epsilon-\Omega_{n-2}-_{\ddots}}},
\nonumber\\
\Delta_n^+
=\frac{I_{n}^2}{\omega\pm{\mbox{i}}\epsilon-\Omega_{n+1}
-\frac{I_{n+1}^2}{\omega\pm{\mbox{i}}\epsilon-\Omega_{n+2}-_{\ddots}}}.
\nonumber
\end{eqnarray}
For any finite period of varying 
of the Hamiltonian parameters 
$\Omega_n$ and $I_n$  
the continued fractions in Eq. (\ref{003}) become periodic 
and can be evaluated by solving quadratic equations. 
As a result  
one gets the exact expressions for $G_{nn}^{\mp}$,
$\langle c_n^+c_n\rangle
=\mp\frac{1}{\pi}\int_{-\infty}^{\infty}
{\mbox{d}}\omega\frac{{\mbox{Im}}G_{nn}^{\mp}}
{{\mbox{e}}^{\beta\omega}+1}$,
and hence for $m_n$ and $\chi_n$. 
Introducing the ``local'' density of states
$\rho_n(\omega)
=\mp\frac{1}{\pi}{\mbox{Im}}G_{nn}^{\mp}$
the formulas for $m_n$ and $\chi_n$ 
can be rewritten as follows
\begin{eqnarray}
\label{004}
m_n=-\frac{1}{2}
\int_{-\infty}^{\infty}{\mbox{d}}\omega
\rho_n(\omega)\tanh\frac{\beta\omega}{2},
\end{eqnarray}
\begin{eqnarray}
\label{005}
\chi_n=-\frac{\beta}{4}
\int_{-\infty}^{\infty}{\mbox{d}}\omega
\frac{\rho_n(\omega)}{\cosh^2\frac{\beta\omega}{2}}.
\end{eqnarray}
Here $\beta=\frac{1}{kT}$ is the reciprocal temperature.
Knowing the local quantities 
$\rho_n(\omega)$, $m_n$ and $\chi_n$ 
one immediately gets 
the (total) density of states 
$\rho(\omega)=\frac{1}{N}\sum_{n=1}^N\rho_n(\omega)$ 
and the (total) magnetization and static susceptibility 
$m=\frac{1}{N}\sum_{n=1}^Nm_n$ and $\chi=\frac{1}{N}\sum_{n=1}^N\chi_n$,
respectively.  

Following the  
procedure outlined above for the 
calculation of $\rho_n(\omega)$ 
(some further details can be found in \cite{010}
where, however, only the total density of states 
$\rho(\omega)$ was considered)
one can easily derive  $\rho_n(\omega)$ 
for chains 
provided the period is not too long. 
For example, for a chain of period 3 
one finds
\begin{eqnarray}
\label{006}
\rho_n(\omega)=
\left\{
\begin{array}{ll}
\frac{1}{\pi}
\frac{\vert{\cal{Y}}_n(\omega)\vert}
{\sqrt{{\cal{C}}(\omega)}},
& {\mbox{if}}\;\;\;{\cal{C}}(\omega)>0,\\
0,
& {\mbox{otherwise}},
\end{array}
\right.
\\
{\cal{Y}}_n(\omega)
=\left(\omega-\Omega_{n+1}\right)\left(\omega-\Omega_{n+2}\right)
-I_{n+1}^2,
\nonumber\\
{\cal{C}}(\omega)=
4I_1^2I_2^2I_3^2
-\left(
I_1^2\left(\omega-\Omega_3\right)
+I_2^2\left(\omega-\Omega_1\right)
+I_3^2\left(\omega-\Omega_2\right)
-\left(\omega-\Omega_1\right)
\left(\omega-\Omega_2\right)
\left(\omega-\Omega_3\right)
\right)^2
\nonumber\\
=-\prod_{j=1}^6\left(\omega-c_j\right),
\nonumber
\end{eqnarray}
where $c_j$ are the six roots of the equation 
${\cal{C}}(\omega)=0$.
For a chain of period 4 
the corresponding result is
\begin{eqnarray}
\label{007}
\rho_n(\omega)=
\left\{
\begin{array}{ll}
\frac{1}{\pi}
\frac{\vert{\cal{W}}_n(\omega)\vert}
{\sqrt{{\cal{D}}(\omega)}},
& {\mbox{if}}\;\;\;{\cal{D}}(\omega)>0,\\
0,
& {\mbox{otherwise}},
\end{array}
\right.
\\
{\cal{W}}_n=
\left(\omega-\Omega_{n+1}\right)
\left(\omega-\Omega_{n+2}\right)
\left(\omega-\Omega_{n+3}\right)
-I_{n+1}^2\left(\omega-\Omega_{n+3}\right)
-I_{n+2}^2\left(\omega-\Omega_{n+1}\right),
\nonumber\\
{\cal{D}}(\omega)=
4I_1^2I_2^2I_3^2I_4^2
-\left(
\left(\omega-\Omega_1\right)\left(\omega-\Omega_2\right)
\left(\omega-\Omega_3\right)\left(\omega-\Omega_4\right)
\right.
\nonumber\\
\left.
-I_1^2\left(\omega-\Omega_3\right)\left(\omega-\Omega_4\right)
-I_2^2\left(\omega-\Omega_1\right)\left(\omega-\Omega_4\right)
\right.
\nonumber\\
\left.
-I_3^2\left(\omega-\Omega_1\right)\left(\omega-\Omega_2\right)
-I_4^2\left(\omega-\Omega_2\right)\left(\omega-\Omega_3\right)
\right.
\nonumber\\
\left.
+I_1^2I_3^2+I_2^2I_4^2
\right)^2
=-\prod_{j=1}^8\left(\omega-d_j\right),
\nonumber
\end{eqnarray}
where $d_j$ are the eight roots of the equation ${\cal{D}}(\omega)=0$.
The analytical calculation of $\rho_n(\omega)$
for chains of larger periods 
can be easily implemented on a small computer
(see the results for periods 6 and 12 presented below).

\section{Magnetization: exact results}

Let us discuss the local magnetic properties 
of the spin model (\ref{001})
induced by regular nonuniformity.
In what follows we restrict ourselves to a case of the uniform field 
$\Omega_n=\Omega$ 
and assume that 
$I_n=I(1+\delta_n)$
where $\delta_n$ is taken 
either in the form
$\delta_n
=\delta\left(\delta_{n,1}-\delta_{n,2}
+\delta_{n,1+p}-\delta_{n,2+p}
+\delta_{n,1+2p}-\delta_{n,2+2p}
+\ldots\;\right)$
or in the form
$\delta_n=-\delta\cos\frac{2\pi n}{p}$.
The parameter $\delta$ 
controls the deviation from uniformity
and the parameter $p$ 
is the period of modulation of the exchange interactions. 
The modulation of exchange interactions 
may be interpreted as a result of the lattice distortion.
For example,
the first ansatz for $\delta_n$ 
corresponds to a displacement of the second site towards the first one, 
of the $(p+2)$th site towards the $(p+1)$th one and so on.
The second ansatz 
corresponds to the displacements of sites  
giving by cosine with period $p$.
Some relevant lattice
configurations for $p=3,4,6,12$ are shown in Fig. 1.
\begin{figure}[ht]
\epsfysize=40mm
\epsfclipon
\centerline{\epsffile{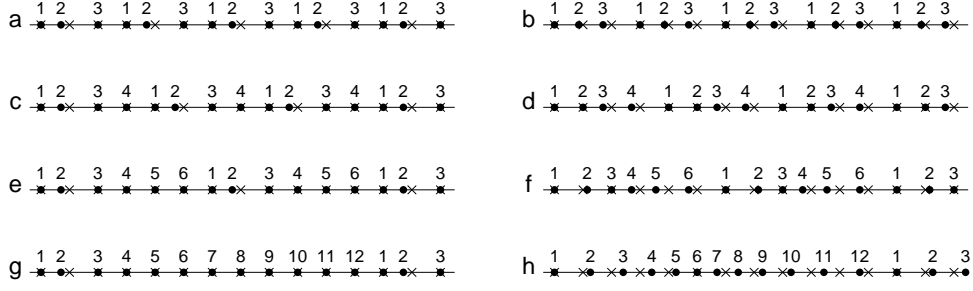}}
\caption[ ]
{\small
Lattice distortions of the form 
$\delta_n
=\delta\left(\delta_{n,1}-\delta_{n,2}
+\delta_{n,1+p}-\delta_{n,2+p}
+\delta_{n,1+2p}-\delta_{n,2+2p}
+\ldots\;\right)$
(a, c, e, g)
and of the form
$\delta_n=-\delta\cos\frac{2\pi n}{p}$
(b, d, f, h) 
for $p=3$ (a, b), $p=4$ (c, d), $p=6$ (e, f) and $p=12$ (g, h).}
\label{fig1}
\end{figure}
In our illustrations presented below we put for concreteness
$I=1$, $\delta=0.6$
(except Figs. 4, 5 where $\delta$ varies from $0.3$ to $0.999$).

In Fig. 2 
\begin{figure}
\epsfysize=200mm
\epsfclipon
\centerline{\epsffile{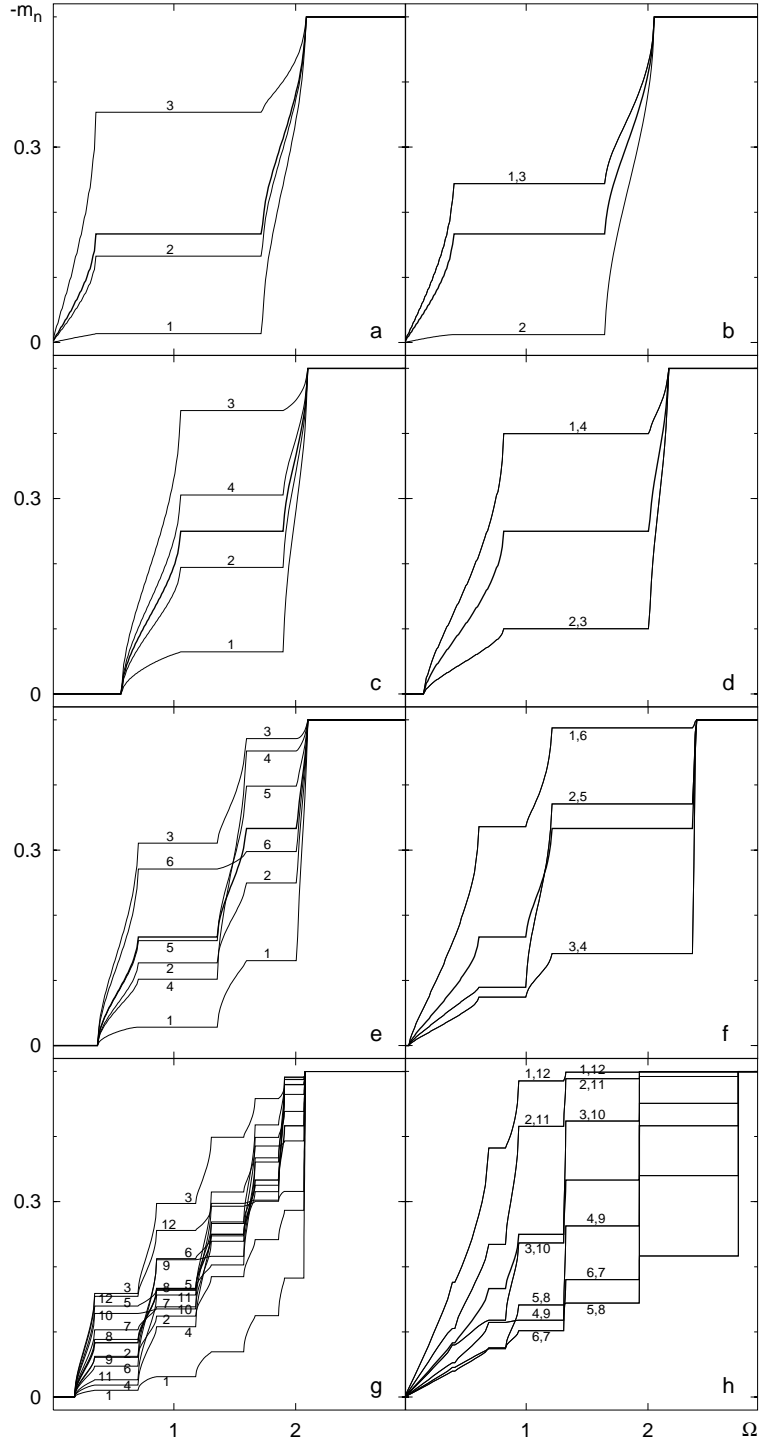}}
\caption[ ]
{\small
Local magnetizations $m_n$ vs. field $\Omega$ 
at zero temperature $\beta=\infty$ 
for the chains shown in Fig. 1 
($I=1$, $\delta=0.6$).
Panel a corresponds to the chain shown in Fig. 1a,
panel b corresponds to the chain shown in Fig. 1b,
and so on;
the curve denoted by $n$ presents the local magnetization $m_n$;
the bold curves present $m$.}
\label{fig2}
\end{figure}
the zero temperature dependences of local magnetizations 
$m_n$ on field $\Omega$ are depicted.  
Similarly to the total magnetization $m$, 
the local magnetizations $m_n$
in regularly alternating $XX$ chains   
exhibit a step--like dependence 
on the applied field. 
From the mathematical point of view 
this is the consequence of a splitting of the initial fermion band 
($\rho_n(\omega)
=\frac{1}{\pi}
\frac{1}
{\sqrt{4I^2-\left(\omega-\Omega\right)^2}}$ 
if 
$\Omega-2\vert I\vert
<\omega<
\Omega+2\vert I\vert$
and
$\rho_n(\omega)=0$
otherwise)
into several subbands,
the edges of which are determined by the roots of equations 
${\cal{C}}(\omega)=0$ (\ref{006}),
${\cal{D}}(\omega)=0$ (\ref{007}), etc..
As a result the zero temperature dependence 
$m_n$ vs. $\Omega$ 
[as well as $m$ vs. $\Omega$] 
must be composed of sharply increasing parts
(as can be seen from Eq. (\ref{004}) 
they appear when 
$\omega=0$ 
remains inside a subband
while $\Omega$ increases) 
separated by the horizontal parts
(they appear when 
$\omega=0$ 
remains outside a subband
while $\Omega$ increases). 
The characteristic fields 
at which the horizontal parts 
of the dependences $m_n$ vs. $\Omega$
begin and end up 
(and therefore the plateau lengths) 
are the same for all $m_n$  
and coincide with the values of such fields for $m$,
since in all the cases
those fields  
are determined by the roots of equations 
${\cal{C}}(\omega)=0$,
${\cal{D}}(\omega)=0$, etc..
However, 
the heights of plateaus $m_n$
are essentially site--dependent quantities. 
Moreover, the values of $m_n$ depend on $\delta$. 
This is in contrast to the behavior of $m$,
since the possible values of $m$ 
are universal and do not depend on 
details of the intersite exchange interactions.
As can be seen, 
for example, in Fig. 2b 
for the lattice shown in Fig. 1b 
the magnetization at site 2 
in weak and in moderate fields 
is always smaller than the magnetizations at sites 1 and 3
and its value depend on $\delta$
(compare the results for $\delta=0.6$ seen in Figs. 2a, 2b, 2e
and for $\delta=1$ presented below).
On the other hand, the sequence $m_{n_1}\le m_{n_2}\le\ldots\le m_{n_p}$ 
may depend on the value of the applied field $\Omega$
as can be seen in Fig. 2e (and Figs. 2g, 2h) .
For example, 
for the chain shown in Fig. 1e
$-m_1<-m_4<-m_2<-m_5<-m_6<-m_3$ 
for $0.7037\ldots\le\Omega\le 1.3578\ldots$ 
but
$-m_1<-m_2<-m_6<-m_5<-m_4<-m_3$  
for $1.5884\ldots\le\Omega\le 2.0139\ldots\;$.
The mentioned property is also visible in Fig. 3 
where the zero temperature magnetization profiles along chains
for various fields
are displayed.
\begin{figure}
\epsfysize=200mm
\epsfclipon
\centerline{\epsffile{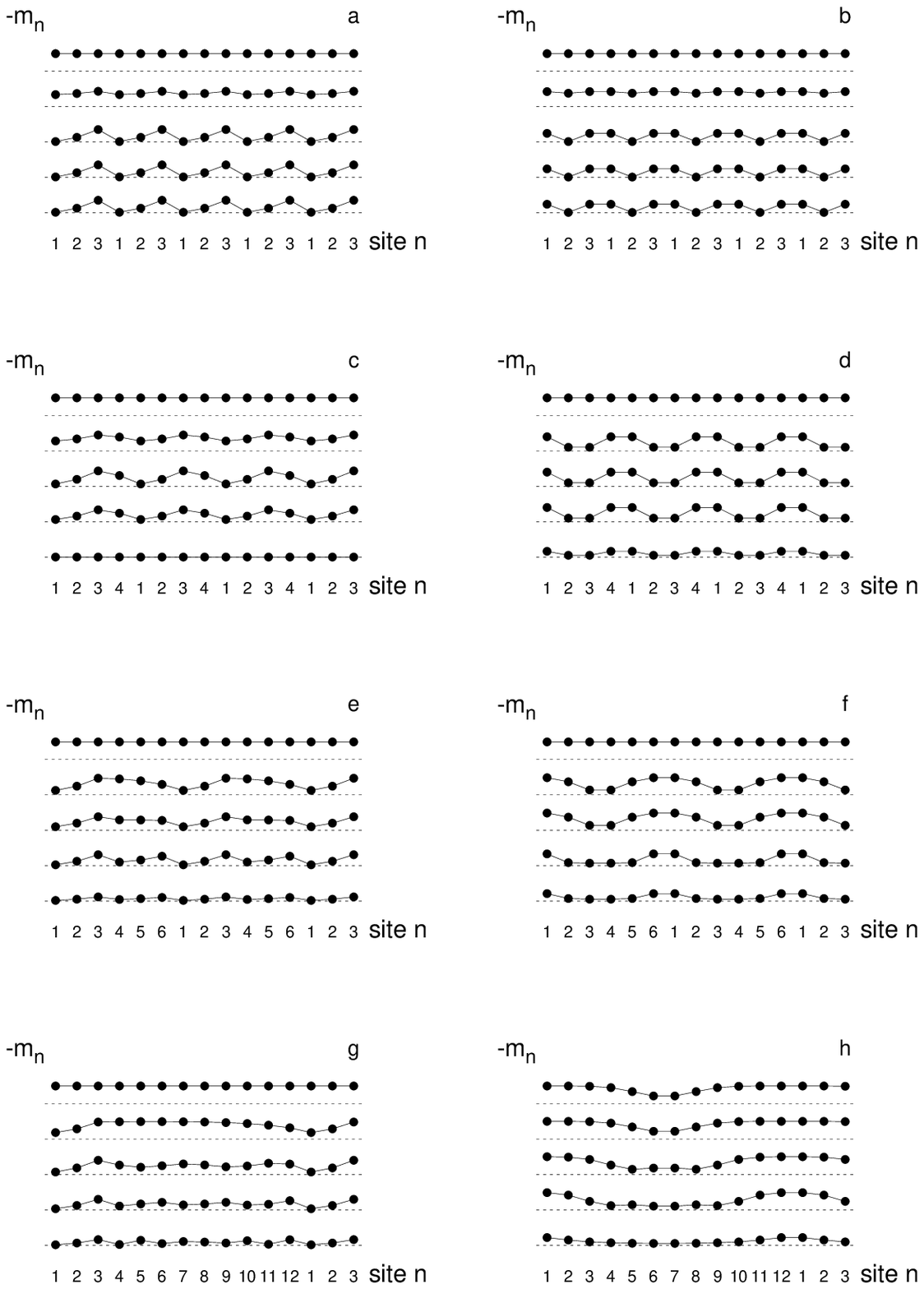}}
\caption[ ]
{\small
Zero--temperature 
magnetization profiles 
$\{m_n\}$ 
at 
$\Omega=0.5, 1, 1.5, 2, 2.5$ (from bottom to top)
for the chains shown in Fig. 1
($I=1$, $\delta=0.6$).
Dashed lines indicate $m_n=0$.
Panel a corresponds to the chain shown in Fig. 1a,
panel b corresponds to the chain shown in Fig. 1b,
and so on.}
\label{fig3}
\end{figure}
Moreover,
the sequence 
$m_{n_1}\le m_{n_2}\le\ldots\le m_{n_p}$ 
for a certain $\Omega$ may depend on $\delta$ 
as can be seen in Fig. 4.
\begin{figure}
\epsfysize=200mm
\epsfclipon
\centerline{\epsffile{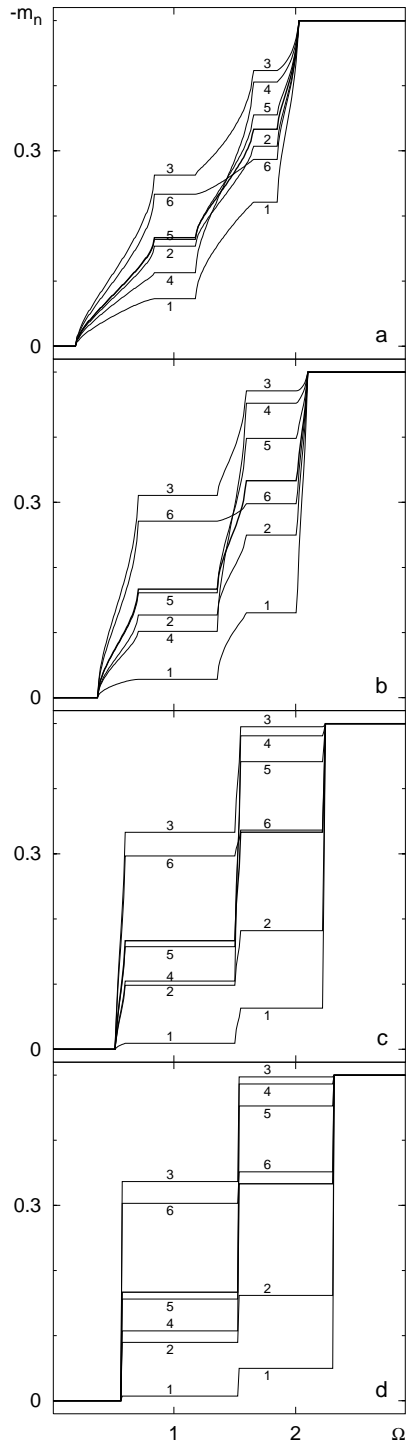}}
\caption[ ]
{\small
Local magnetizations $m_n$ vs. field $\Omega$ 
at zero temperature 
for the chain shown in Fig. 1e
for $\delta=0.3$ (a),
0.6 (b),
0.9 (c),
0.999 (d)
($I=1$).}
\label{fig4}
\end{figure}
The results presented in Fig. 2 (3) 
indicate a relation 
between the periodically modulated exchange interactions 
[or lattice distortion] 
and magnetization profiles at different sites
(magnetization profiles along a chain for various fields), 
which for the  
considered spin-$\frac{1}{2}$ $XX$ chain in a transverse field 
can be traced rigorously.
Such correspondence may be of experimental importance.
Thus, the NMR spectra yield the distribution of the local magnetization
(see, for example, \cite{011,012}). 
In particular, 
such analysis was performed  
to confirm 
that the lattice modulation in the incommensurate phase in CuGeO$_3$  
has the form of a soliton lattice
\cite{011}.

\section{Magnetization: strong--coupling approximation}
 
Although we are able to calculate the magnetization profiles 
for the periodic nonuniform spin-$\frac{1}{2}$ $XX$ chain exactly 
it is worth to consider those results 
from a viewpoint of the  strong--coupling approximation,
which was exploited in a number of papers  
devoted to spin chains \cite{008} and ladders \cite{009}.
On the one hand, by turning to the strong--coupling limit,
we can understand better the magnetization processes 
at zero temperature $\beta=\infty$ 
having a physical picture 
in terms of spins rather than in terms of Jordan--Wigner fermions. 
On the other hand, 
we can use the exact results as a testing ground 
to see how the approximate approach works.
In the strong--coupling limit we put $\delta=1$.
In such a limiting case the chain of period $p$ 
splits into noninteracting clusters 
each consisting of $p$ sites. 
The zero temperature local magnetization in such limit is given by
$m_n=\langle GS\vert s_n^z\vert GS\rangle$
where $\vert GS\rangle$ 
is the ground state eigenvector of the cluster Hamiltonian.
The magnetization plateaus 
arise due to a change of the ground state 
with varying of the field. 

Consider, for example,  
the chain of $p=3$ shown in Fig. 1a.
The relevant 3--site cluster Hamiltonian
has the following eigenvectors and eigenvalues
\begin{eqnarray}
\label{008}
\vert 1\rangle
=\frac{1}{\sqrt{10}}
\vert \downarrow_1\downarrow_2\uparrow_3\rangle
+\frac{2}{\sqrt{10}}
\vert \downarrow_1\uparrow_2\downarrow_3\rangle
-\frac{1}{\sqrt{2}}
\vert \uparrow_1\downarrow_2\downarrow_3\rangle,
\;\;\;\;\;
\vert 2\rangle
=\frac{1}{\sqrt{10}}
\vert \uparrow_1\uparrow_2\downarrow_3\rangle
+\frac{2}{\sqrt{10}}
\vert \uparrow_1\downarrow_2\uparrow_3\rangle
-\frac{1}{\sqrt{2}}
\vert \downarrow_1\uparrow_2\uparrow_3\rangle,
\nonumber\\
\vert 3\rangle
=\vert \downarrow_1\downarrow_2\downarrow_3\rangle,
\;\;\;\;\;
\vert 4\rangle
=\frac{2}{\sqrt{5}}
\vert \downarrow_1\downarrow_2\uparrow_3\rangle
-\frac{1}{\sqrt{5}}
\vert \downarrow_1\uparrow_2\downarrow_3\rangle,
\nonumber\\
\vert 5\rangle
=\frac{2}{\sqrt{5}}
\vert \uparrow_1\uparrow_2\downarrow_3\rangle
-\frac{1}{\sqrt{5}}
\vert \uparrow_1\downarrow_2\uparrow_3\rangle,
\;\;\;\;\;
\vert 6\rangle
=\vert \uparrow_1\uparrow_2\uparrow_3\rangle,
\nonumber\\
\vert 7\rangle
=\frac{1}{\sqrt{10}}
\vert \downarrow_1\downarrow_2\uparrow_3\rangle
+\frac{2}{\sqrt{10}}
\vert \downarrow_1\uparrow_2\downarrow_3\rangle
+\frac{1}{\sqrt{2}}
\vert \uparrow_1\downarrow_2\downarrow_3\rangle,
\;\;\;\;\;
\vert 8\rangle
=\frac{1}{\sqrt{10}}
\vert \uparrow_1\uparrow_2\downarrow_3\rangle
+\frac{2}{\sqrt{10}}
\vert \uparrow_1\downarrow_2\uparrow_3\rangle
+\frac{1}{\sqrt{2}}
\vert \downarrow_1\uparrow_2\uparrow_3\rangle,
\end{eqnarray}
\begin{eqnarray}
\label{009}
E_1=-\frac{1}{2}\Omega-\sqrt{5}I,
\;\;\;
E_2=\frac{1}{2}\Omega-\sqrt{5}I,
\;\;\;
E_3=-\frac{3}{2}\Omega,
\;\;\;
E_4=-\frac{1}{2}\Omega,
\nonumber\\
E_5=\frac{1}{2}\Omega,
\;\;\;
E_6=\frac{3}{2}\Omega,
\;\;\;
E_7=-\frac{1}{2}\Omega+\sqrt{5}I,
\;\;\;
E_8=\frac{1}{2}\Omega+\sqrt{5}I,
\end{eqnarray}
respectively.
As follows from Eq. (\ref{009})
for $0<\Omega<\sqrt{5}I$ the ground state is $\vert 1\rangle$ 
and therefore according to Eq. (\ref{008})
$m_1=0$,
$m_2=-\frac{1}{10}$,
$m_3=-\frac{2}{5}$,
and
$m=-\frac{1}{6}$,
that agreeing with the results for $\delta=0.6$ 
plotted in Fig. 2a.
If $\Omega$ exceeds $\sqrt{5}I$ 
then the ground state is $\vert 3\rangle$ 
and according to Eq. (\ref{008}) one gets
$m_1=m_2=m_3=m=-\frac{1}{2}$.
Similar arguments applied to the chain shown in Fig. 1b yield 
$m_1=m_3=-\frac{1}{4}$, $m_2=0$
if $0<\Omega<\frac{3}{\sqrt{2}}I$
and
$m_1=m_2=m_3=-\frac{1}{2}$
if $\Omega>\frac{3}{\sqrt{2}}I$.

Let us consider further the chain of $p=6$ shown in Fig. 1e.
The eigenvalues and the eigenvectors 
of the relevant 6--site cluster Hamiltonian 
can be easily found numerically.
Thus, the eigenvector 
$\vert S^z=-1\rangle$, 
i.e. with the expectation value of $S^z= s^z_1+\ldots+s^z_6$ 
equals to $-1$, 
which yields the lowest energy in this sector of $S^z$,
is
\begin{eqnarray}
\label{010}
\vert S^z=-1\rangle=
-0.16
\vert \downarrow_1\downarrow_2\downarrow_3
\downarrow_4\uparrow_5\uparrow_6\rangle
+0.21
\vert \downarrow_1\downarrow_2\downarrow_3
\uparrow_4\downarrow_5\uparrow_6\rangle
-0.14
\vert \downarrow_1\downarrow_2\uparrow_3
\downarrow_4\downarrow_5\uparrow_6\rangle
\nonumber\\
+0.23
\vert \downarrow_1\uparrow_2\downarrow_3
\downarrow_4\downarrow_5\uparrow_6\rangle
-0.23
\vert \uparrow_1\downarrow_2\downarrow_3
\downarrow_4\downarrow_5\uparrow_6\rangle
-0.07
\vert \downarrow_1\downarrow_2\downarrow_3
\uparrow_4\uparrow_5\downarrow_6\rangle
\nonumber\\
+0.06
\vert \downarrow_1\downarrow_2\uparrow_3
\downarrow_4\uparrow_5\downarrow_6\rangle
-0.38
\vert \downarrow_1\uparrow_2\downarrow_3
\downarrow_4\uparrow_5\downarrow_6\rangle
+0.41
\vert \uparrow_1\downarrow_2\downarrow_3
\downarrow_4\uparrow_5\downarrow_6\rangle
\nonumber\\
-0.02
\vert \downarrow_1\downarrow_2\uparrow_3
\uparrow_4\downarrow_5\downarrow_6\rangle
+0.39
\vert \downarrow_1\uparrow_2\downarrow_3
\uparrow_4\downarrow_5\downarrow_6\rangle
-0.44
\vert \uparrow_1\downarrow_2\downarrow_3
\uparrow_4\downarrow_5\downarrow_6\rangle
\nonumber\\
-0.25
\vert \downarrow_1\uparrow_2\uparrow_3
\downarrow_4\downarrow_5\downarrow_6\rangle
+0.28
\vert \uparrow_1\downarrow_2\uparrow_3
\downarrow_4\downarrow_5\downarrow_6\rangle
-0.06
\vert \uparrow_1\uparrow_2\downarrow_3
\downarrow_4\downarrow_5\downarrow_6\rangle,
\end{eqnarray}
whereas the eigenvector 
$\vert S^z=-2\rangle$,
i.e. with the expectation value of $S^z$ 
equals to $-2$,
which yields the lowest energy for given $S^z$,
is
\begin{eqnarray}
\label{011}
\vert S^z=-2\rangle=
0.38
\vert \downarrow_1\downarrow_2\downarrow_3
\downarrow_4\downarrow_5\uparrow_6\rangle
-0.22
\vert \downarrow_1\downarrow_2\downarrow_3
\downarrow_4\uparrow_5\downarrow_6\rangle
+0.12
\vert \downarrow_1\downarrow_2\downarrow_3
\uparrow_4\downarrow_5\downarrow_6\rangle
\nonumber\\
-0.05
\vert \downarrow_1\downarrow_2\uparrow_3
\downarrow_4\downarrow_5\downarrow_6\rangle
+0.58
\vert \downarrow_1\uparrow_2\downarrow_3
\downarrow_4\downarrow_5\downarrow_6\rangle
-0.67
\vert \uparrow_1\downarrow_2\downarrow_3
\downarrow_4\downarrow_5\downarrow_6\rangle
\end{eqnarray}
(only two digits after decimal point are preserved 
in the expansion coefficients in Eqs. (\ref{010}), (\ref{011})).
Calculating the expectation values of $s^z_n$
with the help of Eq. (\ref{010}) and Eq. (\ref{011})
one immediately finds that
$\langle S^z=-1\vert s^z_4\vert S^z=-1\rangle
>\langle S^z=-1\vert s^z_6\vert S^z=-1\rangle$
and
$\langle S^z=-2\vert s^z_4\vert S^z=-2\rangle
<\langle S^z=-2\vert s^z_6\vert S^z=-2\rangle$  
just what was observed in Fig. 2e, 
however, for finite $\delta=0.6$.
A validity of the strong--coupling limit predictions  
(which are exact at $\delta=1$)
can be estimated from Fig. 4 
where we plot the exact zero temperature magnetization profiles 
for the chain shown in Fig. 1e
with $\delta$ varying from $0.3$ to $0.999$.

The results of the strong--coupling limit may be used 
to obtain the approximate zero temperature magnetization profiles 
when $\delta$ is slightly less than $1$.
Consider, for example, the chain of $p=3$ shown in Fig 1a.
The Hamiltonian can be naturally split 
into a sum of 3--site cluster Hamiltonians (main part) 
and a part describing the inter--cluster interaction (perturbation).
Only the two lowest levels 
of the 3--site cluster Hamiltonian
should be taken into account 
to describe the zero temperature magnetization profiles
in the region between $m=-\frac{1}{6}$ and  $m=-\frac{1}{2}$.
We assume $\Omega\ge 0$ 
and hence the lowest relevant levels are 
$\vert 1\rangle$ and $\vert 3\rangle$.
Introducing spin $\frac{1}{2}$ operators $\sigma^{\alpha}$
attached to each cluster
which act as
$\sigma^z\vert 1\rangle=-\frac{1}{2}\vert 1\rangle$,
$\sigma^z\vert 3\rangle=\frac{1}{2}\vert 3\rangle$ etc.
one can find that the zero temperature magnetization profiles 
can be calculated with the help of the Hamiltonian of 
the uniform spin-$\frac{1}{2}$ $XX$ chain in a transverse field
\begin{eqnarray}
\label{012}
H=\sum_{l=1}^L
\left(
-\Omega-\frac{1}{2}\sqrt{1+\left(1+\delta\right)^2} I
+\left(\sqrt{1+\left(1+\delta\right)^2} I-\Omega\right)\sigma_l^z
\right)
+\frac{1-\delta^2}{1+\left(1+\delta\right)^2}I
\sum_{l=1}^L
\left(\sigma_l^x\sigma_{l+1}^x+\sigma_l^y\sigma_{l+1}^y\right).
\end{eqnarray}
Moreover, 
the spin operators at sites $s^z_n$ 
are expressed through $\sigma^z$ as follows
\begin{eqnarray}
\label{013}
s^z_1
=-\frac{1}{4}-\frac{1}{2}\sigma^z,
\nonumber\\
s^z_2
=-\frac{2+\left(1+\delta\right)^2}{4\left(1+\left(1+\delta\right)^2\right)}
-\frac{\left(1+\delta\right)^2}{2\left(1+\left(1+\delta\right)^2\right)}
\sigma^z,
\nonumber\\
s^z_3
=-\frac{1+2\left(1+\delta\right)^2}{4\left(1+\left(1+\delta\right)^2\right)}
-\frac{1}{2\left(1+\left(1+\delta\right)^2\right)}\sigma^z.
\end{eqnarray} 
Dotted curves in Fig. 5 show how this effective Hamiltonian works 
while $\delta$ deviates from 1.
\begin{figure}
\epsfysize=200mm
\epsfclipon
\centerline{\epsffile{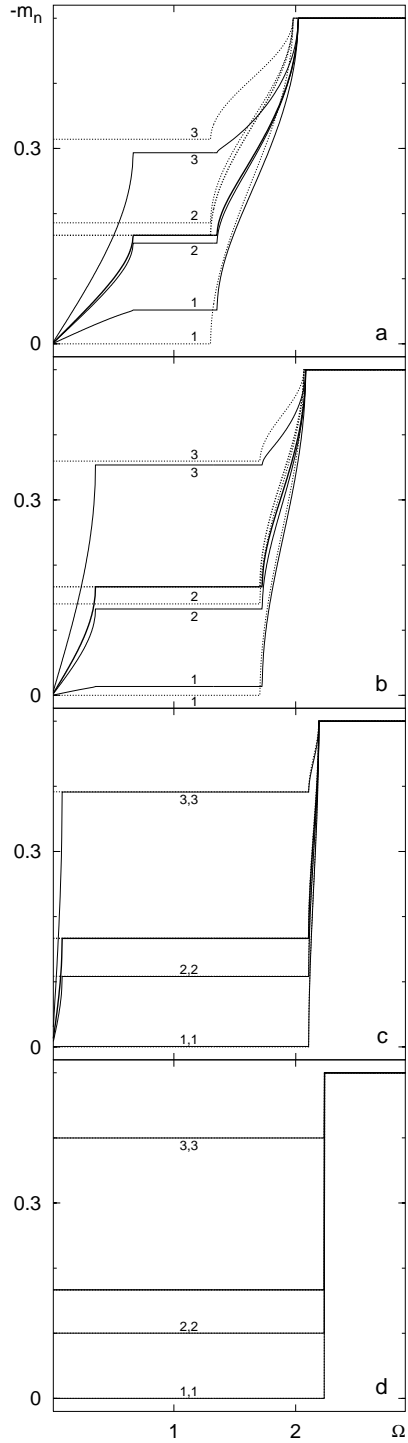}}
\caption[ ]
{\small
Local magnetizations $m_n$ vs. field $\Omega$ at zero temperature 
for the chain shown in Fig. 1a
for $\delta=0.3$ (a), $0.6$ (b), $0.9$ (c), $0.999$ (d) ($I=1$).
Solid curves correspond to the exact results,
dotted curves correspond to the approximate ones 
obtained on the basis of Eqs. (\ref{013}), (\ref{012}).}
\label{fig5}
\end{figure}
We can compare the exact and approximate results.
As $\delta$ decreases the strong--coupling approximation predictions 
for the values of $m_n$, characteristic fields, 
and detailed shape of profiles 
start to differ noticeably from the exact results. 
However, even for $\delta=0.6$ (Fig. 5b) 
the strong--coupling approximation 
yields a reasonably good quantitative picture 
of the magnetization process 
in the region between $m=-\frac{1}{6}$ and $m=-\frac{1}{2}$.
Evidently,
for small $\delta$ (Fig. 5a) 
the strong--coupling approach becomes worse
and fails in the uniform limit $\delta\to 0$ 
as can be seen from Eq. (\ref{013}).
Obviously, the strong--coupling approach 
can be used for other chains with regularly modulated exchange interactions
(e.g. the Heisenberg chain),
for which there are no exact results
and the results presented in Fig. 5 may be of use 
to estimate the accuracy of the strong--coupling approximation.

\section{Susceptibility}

Last let us turn to the local static susceptibilities (\ref{005}).
In Fig. 6 
\begin{figure}
\epsfysize=200mm
\epsfclipon
\centerline{\epsffile{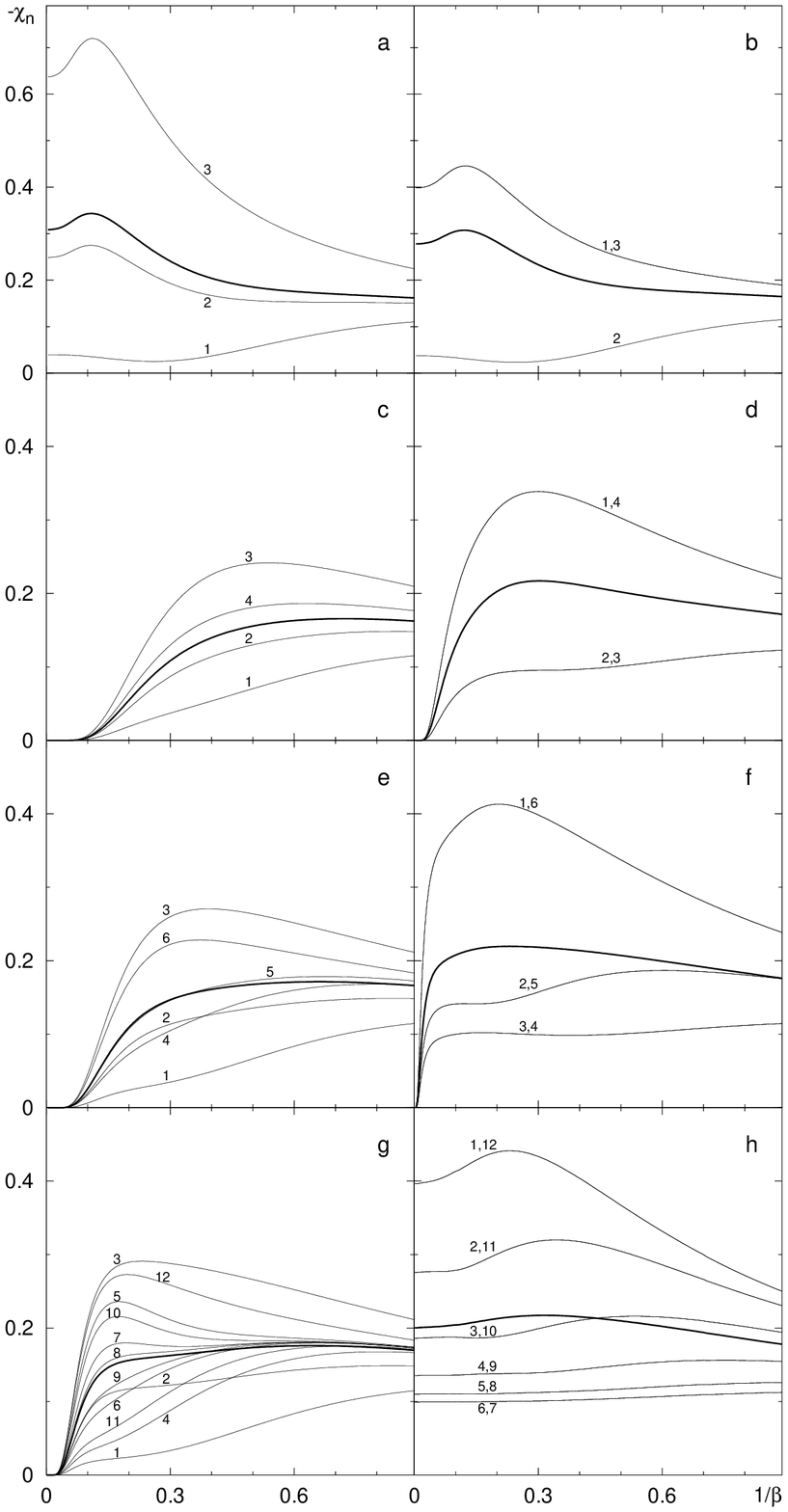}}
\caption[ ]
{\small
Local static susceptibilities $\chi_n$ at $\Omega=0$ vs. temperature 
for the chains shown in Fig. 1
($I=1$, $\delta=0.6$).
Panel a corresponds to a chain shown in Fig. 1a,
panel b corresponds to a chain shown in Fig. 1b,
and so on.}
\label{fig6}
\end{figure}
the temperature dependences of 
$\chi_n$ at $\Omega=0$ are displayed. 
Such dependences are different  
at various sites. 
For example, 
the spin at site 2 of the chain shown in Fig. 1b 
at low temperature 
shows little  
response to the applied field
in contrast to spins at sites 1 and 3
(see Fig. 6b).
It can be also noted 
that some spins in a nonuniform chain 
(e.g., the spins at sites 5 -- 8 for the chain shown in Fig. 1h)
may exhibit 
almost a temperature--independent static susceptibility.
Usually 
for the gapped (gapless) at $\Omega=0$ chains 
shown in Figs. 1c, 1d, 1e, 1f, 1g
(Figs. 1a, 1b, 1h)
the low--temperature behavior of $\chi_n$ and $\chi$ 
differs only in quantitative details.
However, in special cases we can observe a qualitative difference.
Thus, the temperature behavior 
of the static susceptibility for a gapless chain 
mostly increases from a finite value to a maximum 
and then decreases inversely proportionally to temperature,  
whereas, 
for example, the local susceptibility at site 1 (2) 
of the chain shown in Fig. 1a (1b)
exhibits a different behavior:
as the temperature increases 
it decreases achieving a minimum at a finite temperature 
and then increases approaching the high--temperature asymptotic.   

\section{Summary}

To conclude,
we studied 
the local magnetizations and the local static susceptibilities 
of the spin-$\frac{1}{2}$ $XX$ chain in a transverse field 
with regularly alternating exchange interactions. 
These quantities can be calculated exactly 
using the Jordan--Wigner fermionization,
Green function approach 
and continued fractions.
We showed a relation 
between the alteration of the exchange interactions 
and 
i) the zero temperature local magnetizations along the chain
at different external fields
and 
ii) the local susceptibilities along the chain 
at different temperatures. 
We found that the characteristic fields 
at which the zero temperature magnetization plateaus start and end up 
are the same for all sites
contrary to the heights of plateaus,
which are not universal but site--dependent
and which depend on details of intersite interactions 
and the applied field.
We interpreted the observed magnetization profiles 
from a viewpoint of the strong--coupling approach
demonstrating a region of validity of that approximation.
We discussed the temperature behavior of the local susceptibilities.

\vspace{5mm}

The present study was supported by the DFG
(projects 436 UKR 17/2/00 and Ri 615/6--1).
O. D. acknowledges the kind hospitality of the Magdeburg University 
in the summer of 2000 
when this paper was completed.

\end{document}